# Seasonal Evolution on the Nucleus of Comet C/2013 A1 (Siding Spring)

Jian-Yang Li (李荐扬)[1], Nalin H. Samarasinha[1], Michael S.P. Kelley[2], Tony L. Farnham[2], Dennis Bodewits[2], Carey M. Lisse[3], Max J. Mutchler[4], Michael F. A'Hearn[2], W. Alan Delamere[5]

[1] Planetary Science Institute, 1700 E. Ft. Lowell Rd., Suite 106, Tucson, AZ 85719, USA;

jyli@psi.edu, nalin@psi.edu

[2] Department of Astronomy, University of Maryland, College Park, MD 20742, USA;

msk@astro.umd.edu, farnham@astro.umd.edu, dennis@astro.umd.edu, ma@astro.umd.edu

[3] Johns Hopkins University Applied Physics Laboratory, Space Department, 11100 Johns Hopkins Rd., Laurel, MD 20723, USA; carey.lisse@jhuapl.edu

[4] Space Telescope Science Institute, 3700 San Martin Drive, Baltimore, MD 21218-2463, USA;

mutchler@stsci.edu

[5] Delamere Support Service, Bolder, CO 80304, USA; alan@delamere.biz

**Abstract:**

We observed Comet C/Siding Spring using the Hubble Space Telescope (HST) during its close approach to Mars. The high spatial resolution images obtained through the F689M, F775W, and F845M filters reveal the characteristics of the dust coma. The dust production rate of C/Siding Spring, quantified by *Afρ*, is 590±30, 640±30, and 670±30 cm in a 420 km-radius aperture at 38º solar phase angle through the three filters, respectively, consistent with other observations at similar time and geometry, and with model predictions based on earlier measurements. The dust expansion velocity is ~150-250 m s$^{-1}$ for micron-sized dust grains, similar to the speeds found for other comets. The coma has a color slope of (5.5±1.5)%/100 nm between 689 and 845 nm, similar to previous HST measurements at comparable aperture sizes, consistent with the lack of color dependence on heliocentric distance for almost all previously observed active comets. The rotational period of the nucleus of C/Siding Spring is determined from the periodic brightness variation in the coma to be 8.00±0.08 hours, with no excited rotational state detected. The dust coma shows a broad and diffuse fan-shaped feature in the sunward direction, with no temporal morphological variation observed. The projected orientation of the dust feature, combined with the previous analysis of the coma morphology and other characteristics, suggests secular activity evolution of the comet in its inner solar system passage as one previously observed active region turns off whereas new regions exposed to sunlight due to seasonal illumination change.

1. **Introduction**

Dynamically new (DN) comets are those that appear to be on their first return to the inner solar system since being placed in the Oort Cloud over 4 Gyr ago. Compared to returning Oort Cloud comets and short-period comets, DN comets are relatively less evolved and therefore better preserve their physical and chemical states. On the other hand, DN comets are not well characterized, and lack close-up observations from spacecraft. Comet C/2013 A1 (Siding Spring) is a unique DN comet in that it passed by Mars with a closest approach of 140,497 km at 18:28:34 UT on October 19, 2014 (Farnocchia et al. 2015). Such a close approach provides us with an unprecedented opportunity both to image the comet from Mars orbiting spacecraft at close distances, and to study the interactions between a cometary coma and the Martian atmosphere (Schneider et al. 2015, Gurnett et al. 2015, Benna et al. 2015, Gronoff et al. 2014, Yelle et al. 2014). The HiRISE camera onboard the Mars Reconnaissance Orbiter (MRO) imaged the comet at resolutions as high as 140 m/pixel, representing the highest resolution images of a DN comet ever (Farnham et al. 2015, Delamere et al. 2016).

The Hubble Space Telescope (HST) previously observed Comet C/Siding Spring in three epochs (Li et al. 2014) to support the close approach cometary dust impact hazard assessment for Mars orbiters (Tricarico et al. 2014, Kelley et al. 2014, Ye & Hui 2014). Based on the coma morphology, Li et al. (2014) derived two solutions for the spin axis of the nucleus. Later observations (e.g., Opitom et al. 2015, Stevenson et al. 2015), as well as the non-gravitational forces analysis (Farnocchia et al. 2015), provide clues about the long-term activity evolution of C/Siding Spring. Here we report our HST observations of the comet during its Mars close approach, then compare with previous observational results and model predictions, and propose a possible explanation for its long-term behavior. The HST high-resolution view of C/Siding

Spring can also be combined with the HiRISE observations to determine the 3D structure of the cometary coma, and bridge those data to the low-resolution ground-based observations.

2. **Observations and Data Reduction**

Fifteen HST orbits from two observing programs (GO-13675 and GO-13934) were allocated to observe C/Siding Spring with the Wide Field Camera 3 using the UVIS channel (WFC3/UVIS). Six filters were used for the observations: F775W, F689M, F845M, F390M, F547M, and FQ387N. In this report we focus on the dust coma of C/Siding Spring derived from F775W, F689M and F845M filters. The observing geometry and filter characteristics are summarized in Table 1. Note that different field-of-view (FOV) and on-chip binning settings were used for different filters (Table 1). The images collected through the other three filters could contain significant contributions from the CN and $C_2$ gas in the coma, and will be the subject of future work to study the gas in the coma of C/Siding Spring.

For the three filters used to characterize the dust coma, the F845M filter bandpass contains no noticeable emission from the common gas species found in cometary comae, and the F689M filter contains only the weak emission bands of $NH_2$. The F775W filter contains the weak CN (2,0)(3,1) bands near 787 nm. Using the dust/gas ratio, $\log\left(\frac{Af\rho}{Q(CN)}\right) = -22.2$, of C/Siding Spring (Opitom et al. 2015) to scale a spectrum of an extremely gassy comet 122P/de Vico (Cochran & Cochran 2002), we estimate that the gas contamination in those three filters is less than 0.1%.

About 30 images, obtained in eight HST orbits, are impacted by the South Atlantic Anomaly and show a significant increase in cosmic rays. We discarded them from our analysis. In addition, Mars was ~90" from the comet near the closest approach. Although we planned the observations to keep Mars outside of the FOV, four images contain diffuse elevated sky background and sometimes the very bright, structured diffraction spikes of Mars at >20" from the nucleus. When subtracting the sky background, we avoided all artifacts, and measured the sky background from three 200×200 pixel regions in the sunward direction and two sides of the tail near image corners, whenever possible. The sky background appear to be uniform within each region to the level of readout noise, suggesting that our sky measurements should be uncontaminated by the coma. We cleaned all images with resistant-mean filtering leaving the center 20×20 pixel region untouched. Other data reduction, calibration, and photometric measurement follow similar procedures in Li et al. (2013).

## 3. Analysis and Results

*3.1 Dust production rate and outflow speed*

We measured the total brightness of the comet in circular apertures of 1-300 pixels in radius. The statistical uncertainty in the measurement is <0.5%, and the photometric calibration uncertainty of ~5% dominates the total uncertainty. The values of quantity *Afρ* (a proxy for dust production, A'Hearn et al. 1984), averaged over all images used, are listed in Table 1. These values are consistent with the WISE measurement of 720±40 cm in September 2014 acquired at a comparable phase angle of 43º (Stevenson et al. 2015), as well as the MRO/HiRISE measurements of ~600 cm in a 48 km aperture during the encounter (Delamere et al. 2016). Corrected with a factor of 0.37 using the Halley-Marcus comet dust phase function[1] (Marcus 2007; Schleicher & Bair 2011), the *A*(0)*fρ* values (Table 1) suggest a slight decrease in activity since the last previous HST observations in March 2014. To zeroth order, the production rate of micron-sized dust can be modeled as following $r_h^0$ (i.e., constant) in the inner solar system, where $r_h$ is the heliocentric distance, similar to the baseline dust hazard model of Kelley et al. (2014), and much shallower than the typical trend for comets, and $r_h^{-2}$ trend assumed by Tricarico et al. (2014) in their hazard analyses.

---

[1] http://asteroid.lowell.edu/comet/dustphase.html

The standoff distance at which the dust grains emitted in the sunward direction are turned back by solar radiation pressure is determined by the dust grains' size (parameterized by the ratio between radiation pressure force and solar gravity on the dust particle, $\beta$) and ejection velocity ($v$). Using the simplified models of Farnham & Schleicher (2005) and Mueller et al. (2013), the relation $v^2/\beta$ is derived from the observed standoff distance of 6300-9500 km in the F689M and F845M images to be 0.023-0.035 km$^2$ s$^{-2}$, which correspond to velocities 150-200 m s$^{-1}$ for $\beta$=1 and 15-20 m s$^{-1}$ for $\beta$=0.01. These velocities are in good agreement with the pre-encounter coma models of C/Siding Spring (Tricarico et al. 2014, Kelley et al. 2014), which predicted $v$=150-250 m s$^{-1}$ for $\beta$=1, and 16-50 m s$^{-1}$ for $\beta$ =0.01 during the encounter. They are also consistent with the dust velocity trend for many other comets (e.g., Lisse et al. 1998, Krasnopolsky et al. 1987, Sekanina 1982). The dust emission rate of C/Siding Spring therefore appears to be typical.

*3.2 Coma Color*

The azimuthally averaged color slope of the dust coma was measured in concentric annular apertures centered on the nucleus from the F689M and F845M filter images, averaged over all images (Fig. 1). The dust coma within about 14,000 km of the nucleus has a color slope (5.5±0.4)%/100 nm between 688 and 844 nm, with no significant gradient with cometocentric distance. The uncertainty is estimated based on the flux uncertainties at the two wavelengths. Small variations of ~1%/100 nm in in our measurement are likely due to the wings of the point spread functions (PSF) of background stars that are not completely cleaned, suggesting that the uncertainty in our color measurement is more likely 1.5%/100 nm. The color of C/Siding Spring falls into the typical range of active comets (Jewitt 2015). The measured color slope is equivalent to 6.4%/100 nm if we extrapolate the corresponding linear spectrum to the wavelengths of the F438W and F606W filters used in the previous HST measurements (Li et al. 2014). The color of C/Siding Spring's coma at 1.4 AU from the Sun in October 2014 is comparable to that at 4.6-3.7 AU in October 2013 to January 2014, but slightly bluer than the color at 3.3 AU in March 2014, although the difference may not be statistically significant (Li et al., 2014). The nearly constant color of C/Siding Spring is consistent with other observations that suggest no general color trend with heliocentric distance for comets, although most observations were performed at larger heliocentric distances than our observations (Jewitt & Meech 1988, Jewitt 2015). On the other hand, our color slope is much shallower than the ground-based measurement of 15%/100 nm within 10,000 km apertures between BC (445 nm) and RC (713 nm) HB filters (Opitom et al. 2015), although a similar color slope of 5.5%/100 nm between the broadband *I*- and *R*-filters, which have similar equivalent wavelengths to those of the filters we used, are also noticed (Jehin 2015, private communication). The reason is unclear.

*3.3 Rotational Period*

The photometric measurements of C/Siding Spring from our HST images show clear periodic variations (Fig. 2a). We determined the rotational period of C/Siding Spring from the 0".24-radius aperture photometric data. The aperture of this size is minimally affected by sky and background stars, and includes the majority of the PSF (1.7-1.9 pixels FWHM, Dressel 2015), while containing enough pixels for reliable measurements even in the binned F775W images. The measurements from F689M and F845M filters are corrected to F775W filter using a solar spectrum (ASTM-490 2006). We tested the results with and without reddening the solar spectrum by the comet's color, and found a difference of ~0.02 hours in periods, suggesting that the effect of coma color is negligible in period determination. We used the phase dispersion minimization algorithm (PDM, Stellingwerf, 1978) and the ANOVA technique (Schwarzenberg-Czerny, 1996) to search for periods, found consistent results. Fig. 2b shows the PDM result. The primary minimum corresponds to a fundamental frequency of 3.00±0.03/day, or a single-peak period of 8.00±0.08 hours. The uncertainty is estimated using a Monte Carlo approach assuming an individual measurement uncertainty of 0.05 mag for all data points. All other frequencies in the periodogram are harmonics associated with the fundamental frequency, suggesting that the nucleus is in a simple rotational state rather than an excited state. As a DN comet, C/Siding Spring should have spent a much longer time in the outer solar system inactive than the possible rotational damping time scale, which is estimated to be <1/10 of the age of the solar system for a 1-km nucleus rotating at 8-hr period based on Burns & Safronov (1973) and Sharma et al. (2005). An excited rotational state for C/Siding Spring is therefore not expected upon its first return to the inner solar system. Its unexcited rotational state suggests that any torques introduced by the inbound activity have been too small to excite the rotation.

Because the nucleus of C/Siding Spring is likely less than a few km in diameter (Farnham et al. 2015, Delamere et al. 2016), whereas the expected apparent brightness for a 2-km diameter nucleus is ~20.9 mag, the brightness of the comet in a 0".24-radius aperture (280 km at the comet) is dominated by coma. Fig 2a and 2c show that the lightcurve of C/Siding Spring appears to repeat itself every 8 hours. The periodic change in the coma brightness should be dominated by diurnal variations in the dust emission rate that could result in a single-peaked lightcurve, rather than the change in the illuminated cross-section of an elongated nucleus that would cause a double-peaked lightcurve. Therefore, the rotational period of the nucleus is likely 8.00±0.08 hours. The range of the lightcurve in a 280 km-radius aperture is ~26% peak-to-peak, indicating a strong diurnal modulation of the dust emission from a highly heterogeneous nucleus.

*3.4 Coma Morphology*

We searched for temporal variations in the dust coma morphology in the F689M and F845M images by subtracting the median combination of all images from the corresponding filters. No variations are identified other than the slight brightness variations within the center 3-5 pixels that are associated with the diurnal activity variation. Therefore, we median-combined all images from the same filter to improve the signal-to-noise ratio for morphological analysis. After dividing the median-combined images by a coma model with the canonical 1/ρ brightness falloff (ρ is the cometocentric distance), a broad, diffuse dust feature is visible in the sunward direction in both F689M (Fig. 3) and F845M filter images. This feature has a distinctively different morphology from the two relatively sharp, well-defined dust features in the early HST images in late 2013 and early 2014 (Li et al. 2014).

Of the two dust features in previous, distant HST images ($r_h$>3 AU), the position angle (PA) of the northwestern feature in each of the three epochs point toward a common direction in the sky (Li et al. 2014). Such a common direction and the similar morphologies in three epochs suggested that the northwestern features should be the same dust feature viewed from different aspects. The common direction at R.A., decl. = (295º, +43º) should be a reliable measurement of its direction. Based on the viewing geometry of the comet, the northwestern feature in our previous HST imagery should have appeared at ~22º PA, i.e., northeastern, during the close encounter with Mars. But no feature is evident at this location. Its predicted inclination angle of 17º into the sky plane suggests that projection should not affect the observability of the feature. The lack of such a feature in the encounter images (Fig. 3), and the fact that the direction of the earlier northwestern feature is ~115º away from the Sun (Fig. 4) suggest that the northwestern dust feature in the early HST images must have shut off before the Mars close approach.

In contrast, we suggest that the broad sunward feature visible in our HST images corresponds to the previous southeastern feature, or perhaps a combination of the previous southeastern feature and a newly developed feature or features that originate from an area that had only recently moved into sunlight. First, the orientation of the previous southeastern feature in early HST images cannot be fully determined (Li et al. 2014). But our calculation shows that its possible orientations could have the source region or regions illuminated during the close Mars approach, and the possible PAs of that feature are all due west of the comet, consistent with the observed orientation of the sunward feature. Second, the previous southeastern feature appears to be relatively broader and more diffuse than the northwestern feature, consistent with the morphology of this sunward feature in the encounter images. Third, the southeastern feature did not show variations in morphology between the previous three observations, and Li et al. (2014) suggested that it could be a combination of many small features at low latitude. This is consistent with the lack of temporal variation in the morphology of the sunward feature during the encounter. The strong rotational modulation in dust activity suggested by the lightcurve amplitude implies the existence of an active area at low latitude that moves in and out of sunlight as the nucleus rotates, varying its activity level significantly in a diurnal cycle. The modulated activity from this active area adds on top of the overall continuous, sunward activity, similar to the case of 103P/Hartley 2 (Mueller et al. 2013, Belton 2013), forming the sunward feature. The modulated active area must be sufficiently large in order to support a diffuse feature and a >26% variation in activity level without showing distinct variation in the dust morphology it creates.

## 4. Discussion

The dust emitted by C/Siding Spring looks normal when compared to other comets. The production rate, speed of outflow, and color of the dust are all similar to the ensemble averages found for other comets at 1.5 AU. We hypothesize that even though C/Siding Spring is a DN comet, its surface layers may have had time to evolve to a state similar to that of other shorter period comets.

The coma morphology of C/Siding Spring during the observing epochs of HST (Li et al. 2014), the evolution of the activity (Bodewits et al. 2015; Opitom et al. 2015), and the evolution of non-gravitational force (Farnocchia et al. 2015) provide clues about solar insolation conditions on the comet, and help constrain the spin axis direction of the nucleus. The illumination conditions on the comet for one pole solution at R.A., decl. = (295º, +43º) derived by Li et al. (2014) are shown in Fig. 4, together with the best pole solution based on non-gravitational force (Farnocchia et al. 2015). Those two pole solutions result in a similar seasonal effect. Here we propose a plausible scenario to explain the behavior of dust activity and coma morphology of C/Siding Spring from October 2013 through the end of 2014 as the following.

Before March 2014, a considerable fraction of the dust activity originated from a strong active region near the sunlit pole (defined as north pole hereby), which produced the strong and narrow northwestern dust feature with no diurnal modulation expected, unless the nucleus has a complicated shape where self-shading is significant. Another large but weakly active region, or perhaps several widely distributed small regions near the equator produced the relatively broad and diffuse southeastern feature, showing no diurnal modulation in activity in the early HST observations. Pointing ~50º off the orbital plane, the strong, high-latitude or polar feature in the north produced a strong out-of-plane component of the non-gravitational force towards south as observed in this timeframe (Farnocchia et al. 2015).

From March to September 2014, as the subsolar point moved towards the equator, the polar, northwestern feature weakened, and the low-latitude, southeastern feature strengthened. The overall dust activity slowly decreased (Opitom et al. 2015). The out-of-plane component of the non-gravitational force also reached its maximum and started to decrease during this period.

By early September 2014, the Sun crossed the equator, shutting off the strong active region near the north pole, and causing the abrupt decrease in the overall brightness of the comet (Stevenson et al. 2015). The active equatorial region was expected to increase its activity level. The out-of-plane component of the non-gravitational force inverted its direction (Farnocchia et al. 2015) by this time. Without the relatively strong and persistent activity from the previously sunlit polar region, and probably with new active regions developing at low latitude, the rotation of the nucleus resulted in strong modulation in activity, producing a lightcurve with an amplitude of ~26%. In addition, as the Sun moved towards the southern hemisphere, the area that was in polar darkness during pre-perihelion was exposed to strong sunlight, triggering strong and irregular activity, such as the large outburst observed shortly after perihelion (Opitom et al. 2015).

Finally, the Sun reached the solstice point and started to return towards the equator near the end of 2014, and the activity level dropped more rapidly since then, as is typical for DN comets (Whipple 1978, A'Hearn et al. 1995).

This research is supported by NASA through grants HST-GO-13675 and HST-GO-13934 from the STScI. This research made use of Astropy (Astropy Collaboration et al. 2013).

**Figure Captions:**

**Figure 1.** Color slope of C/Siding Spring's dust coma with respect to cometocentric distance.

**Figure 2.** (a) The lightcurve of C/Siding Spring measured from images dominated by dust. The magnitudes from F689M and F845M images are not adjusted for the coma color in this plot, but the adjustment does not affect the period determination (see text). The solid curve is a best-fit sinusoid using the period derived from the PDM analysis. The close encounter of C/Siding Spring with Mars occurred at 2014-Oct-19.77 UT. (b) The PDM analysis showing the $\Theta$ statistic with respect to frequency. The minimum dispersion is at 3.00/day frequency. (c) The phased lightcurves for a single-peaked lightcurve (period 8 hours) and double-peaked lightcurve (period of 16 hours), respectively.

**Figure 3.** The median-combined and $1/\rho$ ratioed image of C/Siding Spring in the F689M filter. North is up and east to the left. The FOV is about 20,000 km for the enhanced image, and about 34,000×29,000 km for the inset. The direction of the Sun, orbital velocity vector, and the projected direction of the northeastern feature in previous HST images are marked by arrows. A prominent diffuse feature is in the sunward direction. The median-combined image before $1/\rho$ enhancement is in the inset with a logarithmic brightness scale.

**Figure 4.** The subsolar latitudes for C/Siding Spring for pole at R.A., decl. = (295º, +43º) from Li et al. (2014) (solid line) and (243º, -14º) from Farnocchia et al. (2015) (dashed line). The three vertical dash-dot lines mark the epochs of the previous HST observations in Li et al.

(2014). The two vertical solid lines mark the close encounter with Mars (2014-10-19T18:28:39) and perihelion (2014-10-25T07:52:39).

**Table 1.** Summary of observing geometry, filter characteristics, data, and dust production rates.

| Start UT | 2014-10-19T01:00:45 | | |
|---|---|---|---|
| End UT | 2014-10-20T14:25:29 | | |
| Heliocentric distance | 1.40 AU | | |
| Geocentric distance | 1.61 – 1.65 AU | | |
| Phase angle | 37.8º – 37.1º | | |
| Native pixel scale at comet | 46.8 – 47.7 km | | |
| Filters | F775W | F689M | F845M |
| Filter effective wavelength (nm) | 764.7 | 687.6 | 843.6 |
| Filter equivalent width (nm) | 117.1 | 68.3 | 78.7 |
| Number of images | 103 | 14 | 14 |
| Field of view | 160"×160" | 80"×80" | 80"×80" |
| Pixel size (arcsec) | 0.12 (binned 3×3) | 0.04 | 0.04 |
| Magnitude[*&+] | 15.70 | 16.35 | 15.94 |
| $A(\alpha)f\rho$[&+] (cm) | 640±30 | 590±30 | 670±30 |
| $A(0)f\rho$[&+] (cm) | 1710 | 1590 | 1790 |

[*] Magnitudes are in Vega-magnitude system based on Kalirai et al. (2009).

[&] Measured from a 0".36-radius (420 km projected at the comet) aperture. Error bars are 1-σ for 5% uncertainty.

[+] Averages over all images (except for those discarded, see text) through the corresponding filter.

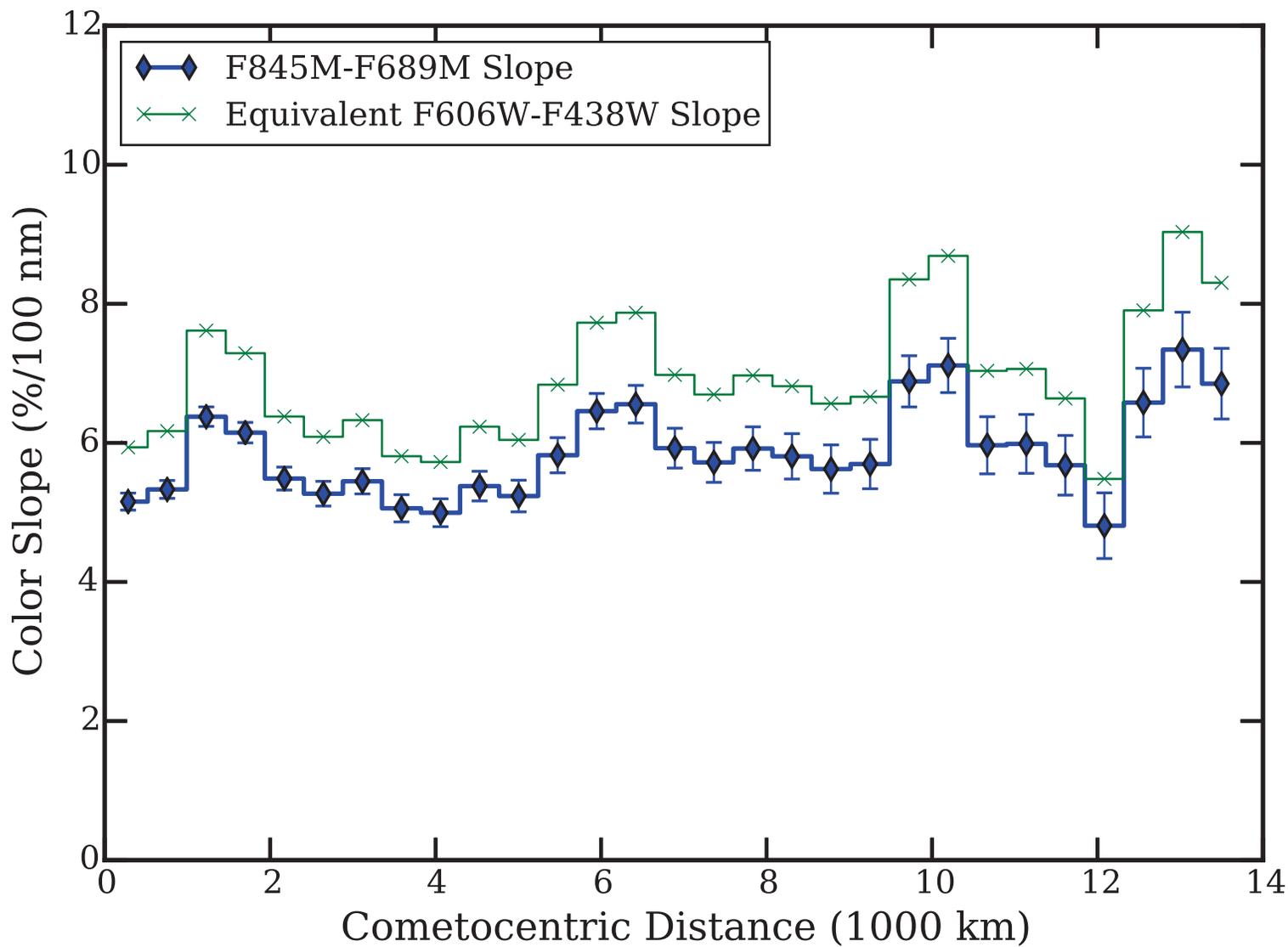

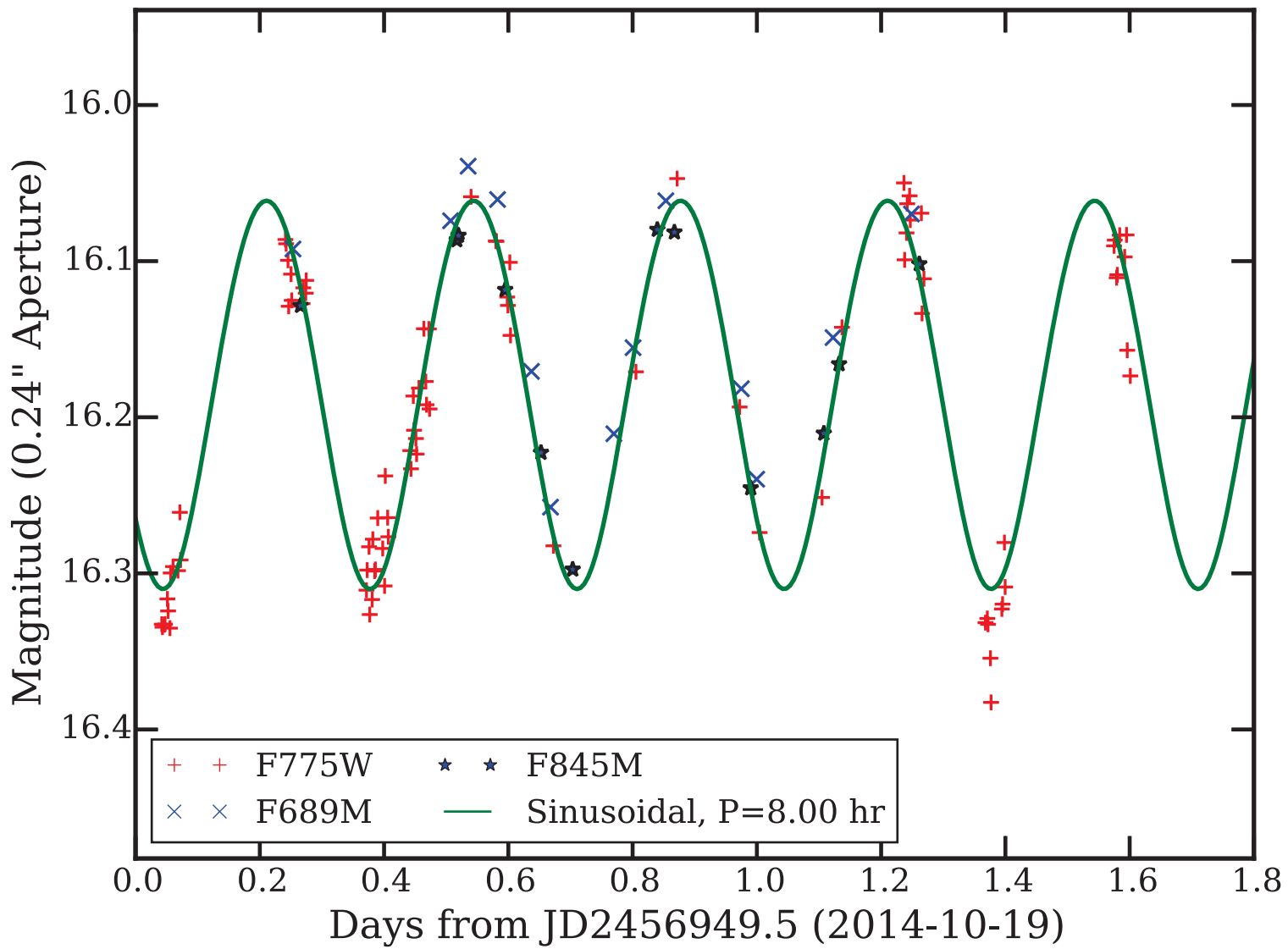

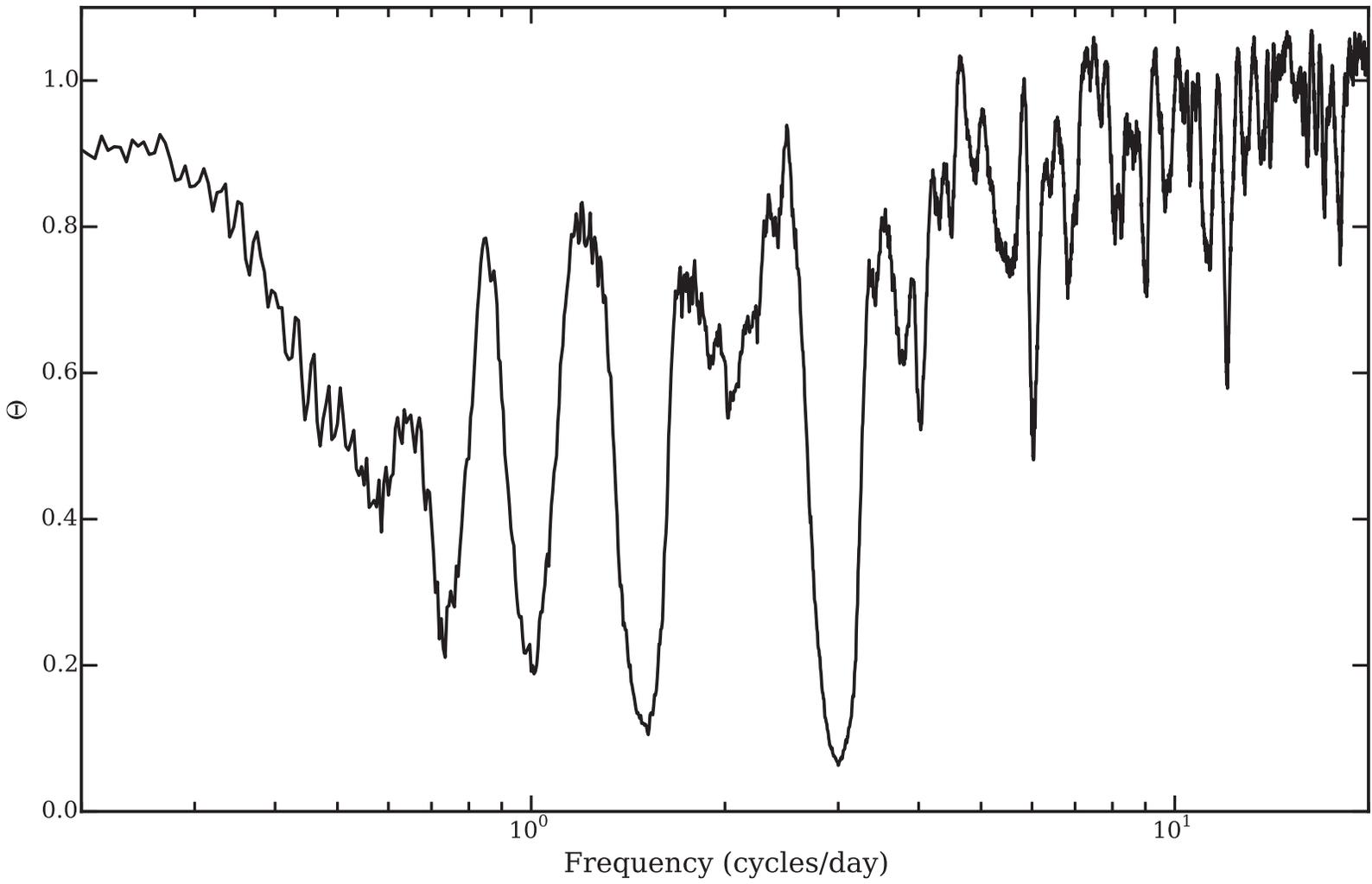

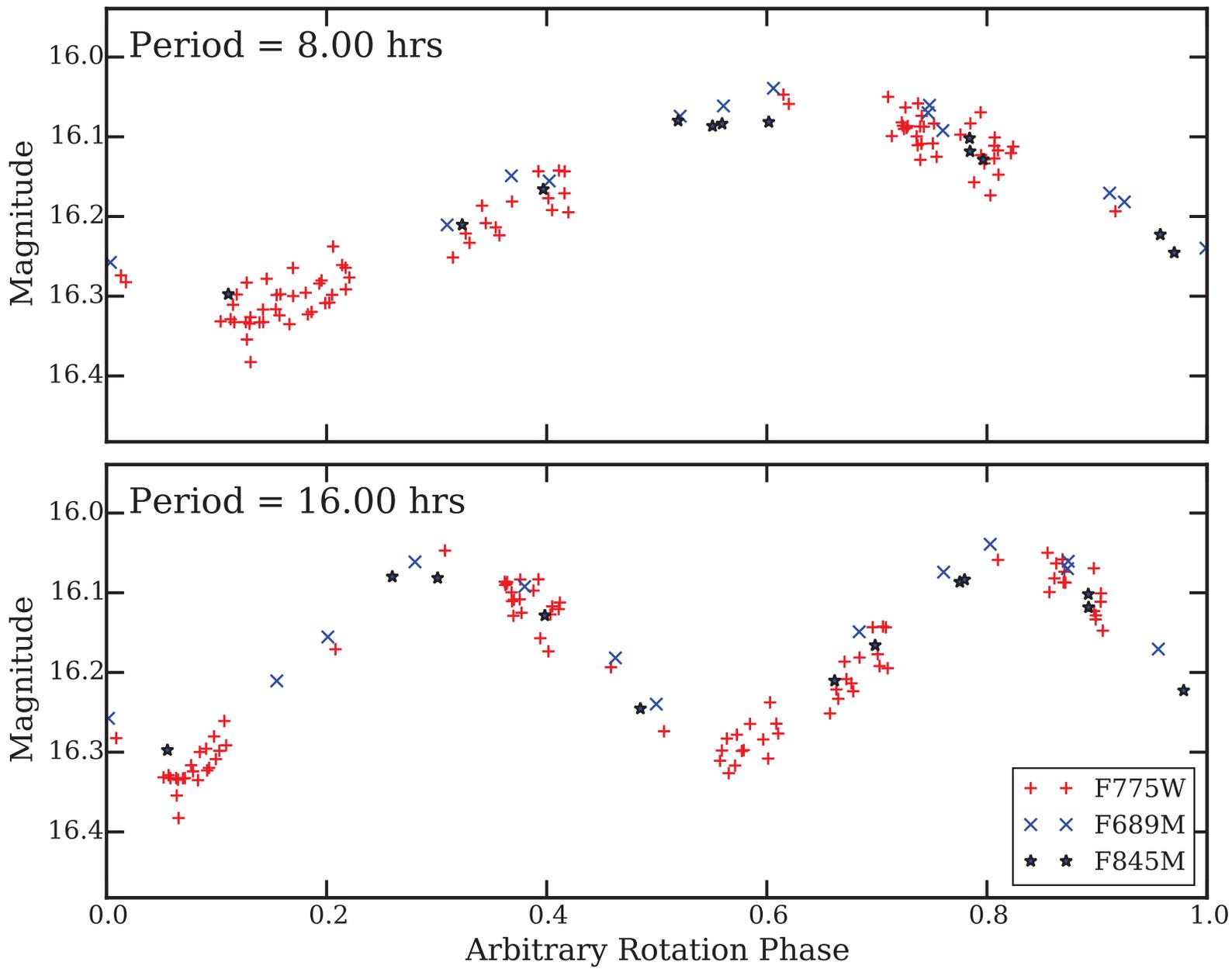

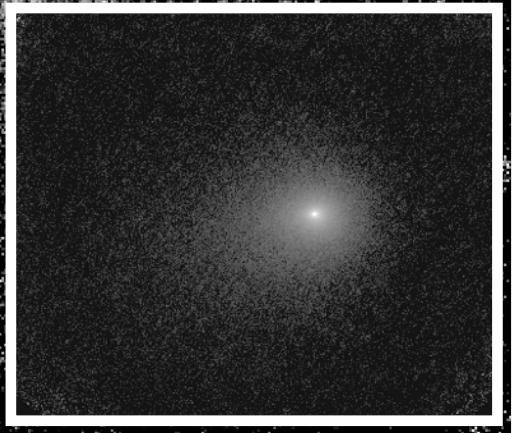

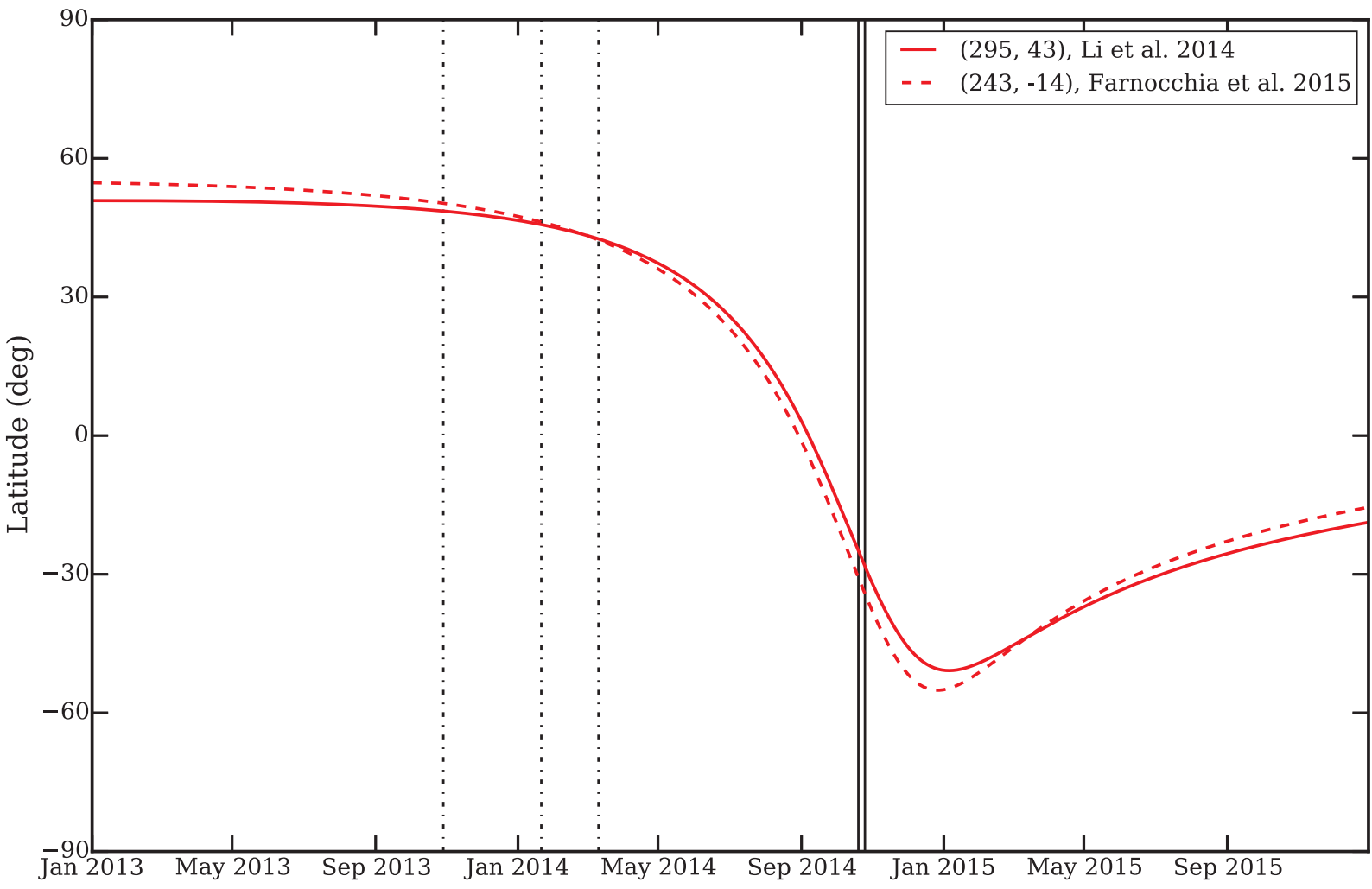